\documentclass[superscriptaddress,aps,prl,twocolumn]{revtex4-1}
\usepackage[hmargin=2.0cm,vmargin=2.3cm]{geometry}
\usepackage{textcomp}
\usepackage{graphicx}
\usepackage{textpos}
\usepackage{amssymb}
\usepackage{amsmath}
\usepackage{hyperref}
\usepackage{txfonts}
\usepackage{natbib}
\usepackage{etoolbox}
\usepackage{caption}
\newcommand{\ket}[1]{\left|#1\right\rangle}

\newcommand{\Ca}{$^{40}$Ca$^{+}$}
\newcommand{\Ba}{Ba$^{+}$}
\newcommand{\Yb}{Yb$^{+}$}

\begin{document}
\title{Strongly enhanced effects of Lorentz symmetry violation in entangled Yb$^+$ ions\\}

\author{V. A. Dzuba}
\affiliation{School of Physics, University of New South Wales, Sydney 2052, Australia}
\author{V. V. Flambaum}
\affiliation{School of Physics, University of New South Wales, Sydney 2052, Australia}
\author{M. Safronova}
\affiliation{Department of Physics and Astronomy, University of Delaware, Newark, Delaware 19716, USA}
\affiliation{Joint Quantum Institute, National Institute of Standards and Technology and the University of Maryland, College Park, Maryland, 20742, USA}
\author{S. G. Porsev}
\affiliation{Department of Physics and Astronomy, University of Delaware, Newark, Delaware 19716, USA}
\affiliation{Petersburg Nuclear Physics Institute, Gatchina, Leningrad District 188300, Russia}
\author{T. Pruttivarasin}
\affiliation{Quantum Metrology Laboratory, RIKEN, Wako, Saitama 351-0198, Japan}
\author{M. A. Hohensee}
\affiliation{Lawrence Livermore National Laboratory, Livermore, California 94550, USA}
\author{H. H\"affner}
\affiliation{Department of Physics, University of California, Berkeley, California 94720, USA}

\maketitle

\date{\today}
\textbf{
Lorentz symmetry is one of the cornerstones of modern physics. However, a number of theories aiming at unifying gravity with  the other fundamental interactions including string field theory suggest violation of Lorentz symmetry\cite{KosRus11,KosteleckyPottingPRD1995,Horava,Pospelov}.
 While the energy scale of such strongly Lorentz symmetry-violating physics is much higher than that currently attainable by particle accelerators, Lorentz violation may nevertheless be detectable via precision measurements at low energies\cite{KosteleckyPottingPRD1995}.
Here, we carry out a systematic theoretical investigation of the sensitivity of a wide range of atomic systems to violation of local Lorentz invariance (LLI). Aim of these studies is to identify which atom shows the biggest promise to detect violation of Lorentz symmetry.
We identify the Yb$^+$ ion
as an ideal system with high sensitivity as well as excellent experimental controllability.
By applying quantum information inspired technology to Yb$^+$, we expect tests of LLI violating physics in the electron-photon sector to reach  levels of $10^{-23}$, five orders of magnitude more sensitive than the current best bounds  \cite{Pruttivarasin2014,Eisele2009,Hohensee2013c}. Most importantly, the projected sensitivity  of $10^{-23}$ for the Yb$^+$ ion tests will allow for the first time to probe whether Lorentz violation is minimally suppressed at low energies for photons and electrons.}

Formally, we can classify LLI-violating effects in the framework of the Standard Model Extension (SME)\cite{KosRus11}. The SME is an effective field theory that maintains Lorentz invariance of the total action, energy-momentum conservation, and gauge invariance, but
supplements the Standard Model Lagrangian with all combinations of the SM fields that are not term-by-term Lorentz invariant. Here we focus on the $c_{\mu \nu}$ tensor term of the SME Lagrangian signifying the dependency of the maximally attainable velocity of a particle with respect to its propagation direction.
SME allows for a violation of LLI for each type of particle, making it is essential to verify LLI in different systems at a high level of precision. As a result, LLI tests have been conducted  for the photons \cite{Eisele2009},  protons \cite{Wolf2006}, neutrons  \cite{Allmendinger2013,Romalis2011}, electrons
\cite{Hohensee2013c,Pruttivarasin2014}, and neutrinos \cite{MINOS2012} with the detailed summary of all current limits given in \cite{v8}.

\begin{figure}
\includegraphics[width=0.48\textwidth]{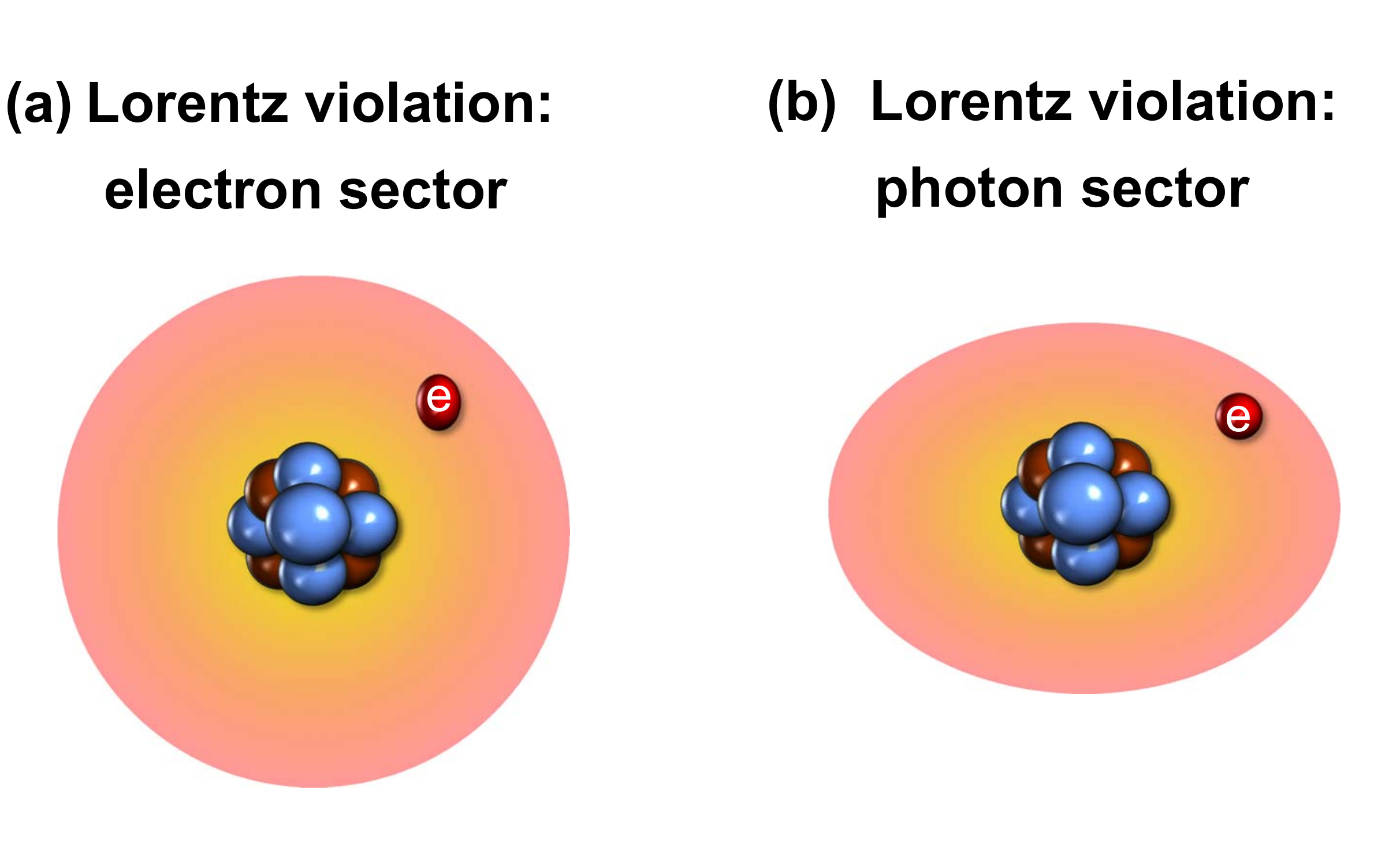}
\caption{
To analyze the atomic LLI experiments, we need to pick a reference frame which allows for two interpretations of the result:
(a) we can either assume that the Coulomb potential  is symmetric and
	any Lorentz-violating (LV) signal is attributed to the electron; or
(b) we assume that electron obeys Lorentz symmetry
and any LV signal is attributed to the photon sector.}
\label{fig1}
\end{figure}

Testing LLI of the electron motion in a Coulomb potential created by a nucleus has the appeal of testing for new physics in a well understood system. In these atomic experiments \cite{Hohensee2013c,Pruttivarasin2014}, one searches for the variations of the atomic energy levels when the orientation of the electronic wave function is rotated with respect to the hypothetical preferred reference frame.
The analysis of any experiment requires the selection of the preferred reference frame. In the present context, this  lends itself to the interpretation  of the results in terms of the Lorentz symmetry test in either electron or photon sectors as illustrated in Fig.~\ref{fig1}.

In the present work, we discuss how to drastically improve LLI tests in the electron-photon sector.
We identify several factors that affect the precision of the LLI test with atomic systems:\\
\noindent \textbf{1.} The lifetime of an atomic state used to probe Lorentz-violating effects.\\
\noindent \textbf{2.} The sensitivity of this state to the LLI-violation effects, i.e., the size of the matrix element of the corresponding operator.
 The LLI violating interaction \cite{Pruttivarasin2014} contains second power of the momentum operator $p$.
Therefore, corresponding matrix elements are expected to be large  for orbitals with large kinetic energy.
This happens for the atomic $4f$-electrons localized deep inside the atom in the area of large potential and kinetic energy in some atomic systems.\\
\noindent \textbf{3.} Good theoretical understanding of the system and availability of already developed experimental techniques.\\

 We find that the metastable $4f^{13} 6s^2$~$^2\text{F}_{7/2}$ state of Yb$^+$ has extremely favorable characteristics such as being very sensitive to LLI violations as well has having an exceptionally long lifetime on the order of 10 years~\cite{Roberts1997}.  In addition, the experimental methods
 for Yb$^+$ are well developed in the context of atomic clocks and measuring the variation of fundamental constants. In particular, the electric-octupole E3 transition between  the $4f^{13} 6s^2$~$^2\text{F}_{7/2}$ excited state and the ground state  \cite{HunLipTam14,GodNisJon14,Huntemann2012} has been studied in detail for these applications. Yb$^+$ ions are also used in quantum information research \cite{IslCamKor13}. As a result, the Yb$^+$ ion is a well-studied system for precision control and manipulation of its atomic states making it particularly well-suited for searches of Lorentz violation signature.

The relevant electronic Lagrangian in the SME (in atomic units) is given by
\begin{align}
\mathcal{L} = \frac{1}{2}i\bar{\psi}(\gamma_\nu+c^{\prime}_{\mu\nu}\gamma^\mu)\stackrel{\leftrightarrow}{D^\nu}\psi-\bar{\psi}\psi, \label{eq:SME_lagrangian}
\end{align}
where $\psi$ is a Dirac spinor, $\gamma^\mu$ are the Dirac matrices, $\bar{\psi}\stackrel{\leftrightarrow}{D^\nu}\psi\equiv\bar{\psi} D^\nu \psi-\psi D^\nu \bar{\psi}$ with $D^\nu$ being the covariant derivative. The tensor $c^{\prime}_{\mu\nu}=c_{\mu\nu}+k_{\mu\nu}/2$ that characterizes the LLI-violation effects
contains Lorentz-violation parameters from both
the electron ($c_{\mu \nu}$) and the photon ($k_{\mu\nu}$) sectors
\cite{KosRus11,KosteleckyPottingPRD1995} as illustrated by Fig.~\ref{fig1}.

From Eq. (\ref{eq:SME_lagrangian}), violations of Lorentz invariance and Einstein's equivalence principle in bound electronic states result in a small shift of the energy levels described by a Hamiltonian\cite{Hohensee2013c}
\begin{equation}
\delta H=-\left(  C_{0}^{(0)}-\frac{2U}{3c^{2}}c_{00}\right)
\frac{\mathbf{p}^{2}}{2}-\frac{1}{6}C_{0}^{(2)}T^{(2)}_{0},\label{eq1}
\end{equation}
where $\mathbf{p}$ is the momentum of a bound electron, $c$ is the speed of light, and U is the Newtonian gravitational potential. The parameters $C_0^{(0)}$, $c_{00}$, and $C_{0}^{(2)}$ are elements in the $c_{\mu \nu}$ tensor. The relativistic form of the $\mathbf{p}^2$ operator is $c\gamma_0\gamma^j p_j$. The nonrelativistic form of the $T^{(2)}_{0}$ operator is
$T^{(2)}_{0}=\mathbf{p}^{2}-3p_{z}^{2},$ and the relativistic form is $T^{(2)}_{0}  = c\gamma_0\left(\gamma^j p_j-3\gamma^3 p_3\right)$, with $z$ (the 3rd spatial component) defined by the quantization axis.

The value of $c_{\mu\nu}$ is specified in the Sun's rest frame. Because of the Earth's motion, time dependent-Lorentz transformations from the Sun's rest frame to the local laboratory frame give rise to time-dependence of the local observables that involve any of the $C_0^{(0)}$, $c_{00}$, and/or $C_{0}^{(2)}$ parameters. The elements $c_{JK}$ in $c_{\mu\nu}$ where $J,K=\{1,2,3\}$ (spatial components), which describe the dependence of the kinetic energy on the direction of the momentum, have a leading order time-modulation period related to the sidereal day (12-hr and 24-hr modulation). Other elements, $c_{TJ}$ (where $T = 0$) and $c_{00}$,  describe the dependence of the kinetic energy on the boost of the laboratory frame, and have the leading order time-modulation period related to the sidereal year.

\begin{figure}
\includegraphics[width=0.48\textwidth]{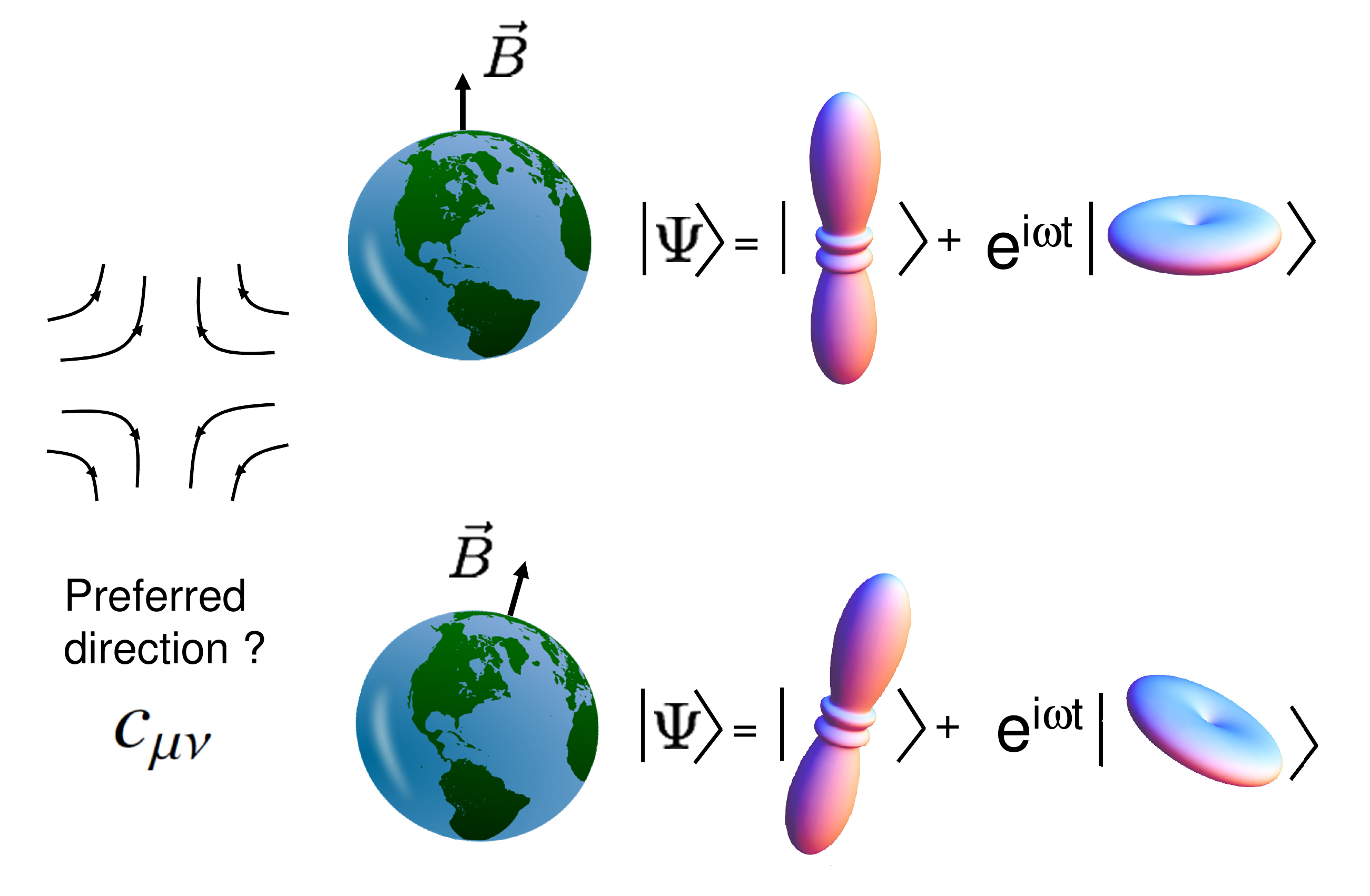}
\caption{ In a magnetic field $\vec{\bf{B}}$, two orthogonal orientations of the electronic wave function
   of the $4f^{13} 6s^2$~$^2F_{7/2}$ $m_J$ manifold in Yb$^+$, with  $m_J=|1/2|$ and $m_J=|7/2|$, will have a different response to the Lorentz-violating effect quantified by the
$c_{\mu \nu}$ tensor. This results in the energy difference between these two states while the Earth rotates
   the wave function with respect to the hypothetical fixed reference frame.}
\label{fig2}
\end{figure}

Yb$^+$ ground state has [Xe]$4f^{14} 6s$ ground state, and is essentially a heavier analog of Ca$^+$  which was used in the most recent experiment \cite{Pruttivarasin2014}, but has additional set of states resulting
from excitation of the filled $4f$ shell. Yb$^+$ also has a metastable excited  $5d_{5/2}$ state analogue to Ca$^+$.
However, the $4f$ shell in Yb$^+$  is localized much deeper in the atom than the $6s$ or
any other valence states, such as $5d$.
As a result, we expect the $4f^{13} 6s^2$~$^2\text{F}_{7/2}$ metastable level of Yb$^+$ to be very sensitive to
the Lorentz-violating effects. Indeed, we find that the LLI sensitive $T^{(2)}_0$ matrix element for this state is over an order of magnitude larger than for the $4f^{14} 5d$ states.
The results of the calculations  are summarized in Table~\ref{tab1}, where we also list $T^{(2)}$ matrix elements
  for $nd$ states of Ca$^+$, Ba$^+$, and Yb$^+$, which all have similar values.

\begin{table}[ht]
\caption{\label{tab1} Reduced matrix elements of $T^{(2)}$ operator in
 Ca$^+$, Ba$^+$, Yb$^+$ ions in atomic units. Ca$^+$ values are from Ref.~\cite{Pruttivarasin2014}.  }
\begin{ruledtabular}
\begin{tabular}{llc}
Ion    & State &    $\langle J||T^{(2)}||J \rangle$       \\ \hline
Ca$^+$ & $3d$~$^2D_{3/2}$ &   7.09(12) \\
       & $3d$~$^2D_{5/2}$ &  9.25(15)   \\ [0.4pc]
Ba$^+$ & $5d$~$^2D_{3/2}$ & 6.83   \\
       & $5d$~$^2D_{5/2}$ & 8.65   \\ [0.4pc]
Yb$^+$ & $4f^{14} 5d$~$^2D_{3/2}$ &  9.96 \\ [0.4pc]
       & $4f^{14} 5d$~$^2D_{5/2}$ &   12.08   \\ [0.4pc]
       & $4f^{13} 6s^2$~$^2F_{7/2}$ &  -135.2 \\
\end{tabular}
\end{ruledtabular}
\end{table}

The calculations of the $T^{(2)}$ reduced matrix elements for the $nd$ states were carried out using the all-order (linearized coupled-cluster) method~\cite{SafJoh08} as described in the Methods section.

In the case of Ca$^+$ ion, the frequency shift between the $m=5/2$ and $m=1/2$ $3d~^2D_{5/2}$ states is given by \cite{Pruttivarasin2014}:
\begin{equation}
\frac{1}{h}\left( E_{m_J=5/2} -E_{m_j=1/2}\right) =-4.45(9)\times10^{15} ~{\rm Hz} \cdot C_{0}^{(2)}.
\end{equation}
In the case of the $4f^{13} 6s^2$~$^2F_{7/2}$  state in \Yb, the frequency shift between the $m=7/2$ and $m=1/2$ states is
\begin{equation}
\frac{1}{h}\left( E_{m_J=7/2} -E_{m_J=1/2}\right) = 6.14\times10^{16} ~ {\rm Hz} \cdot C_{0}^{(2)},
\label{eq:yb_sensitivity}
\end{equation}
giving factor of 15 enhancement in comparison to Ca$^+$ due to the increased $T^{(2)}$ matrix element.
A further significant advantage as compared to Ca$^+$ arises from much longer lifetime of this state as will be discussed below.

 We discuss now to detect potential LLI violations by measuring the energy difference between the $m_J=|7/2|$ and $m_J=|1/2|$
 $4f^{13} 6s^2$~$^2F_{7/2}$  states in Yb$^+$
  (see Fig.~\ref{tab2}).

Typically, the main source of noise in such measurements is due to magnetic field fluctuations. In order to remove this noise, the work of
 Ref.~\cite{Pruttivarasin2014} used a superposition of two ions with a state prepared in a decoherence-free subspace (DFS) \cite{Roos2006,Chwalla2007}.
 To implement the DFS technique with two trapped Yb$^+$ ions, one will monitor   the phase evolution difference of the state $\ket{\Psi} = \frac{1}{\sqrt{2}}(\ket{1/2, -1/2} + \ket{7/2, -7/2})$, where $\ket{m_1, m_2}$ represents the state with $m_J=m_1$ and $m_2$ for the first and second ion, respectively, in the F$_{7/2}$ manifold. The target state $\ket{\Psi}$ can be prepared by creating a product state, $\ket{\Psi^\text{P}} = \frac{1}{2}\left(\ket{-1/2}+\ket{-7/2}\right)\otimes(\ket{+1/2}+\ket{+7/2})$, which dephases into a mixed state that contains $\ket{\Psi}$ with 50\% probability. A direct preparation of an entangled state would result in an increase of the signal-to-noise ratio by a factor of two, however, adds significant experimental complications.

\begin{figure}
\includegraphics[width=0.48\textwidth]{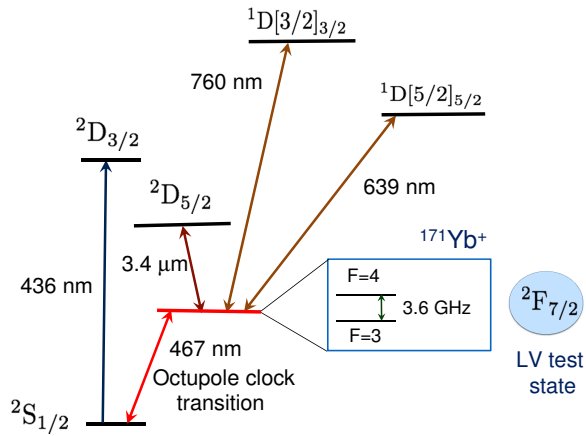}
\caption{Relevant energy levels of \Yb~ with the hyperfine splitting shown for the $4f^{13}6s^2$~$^2F_{7/2}$ state of the $^{171}$\Yb~isotope.}
\label{fig:scheme}
\end{figure}

Starting from the ground state of \Yb, the state $\ket{\Psi} = \frac{1}{\sqrt{2}}(\ket{1/2,-1/2} + \ket{7/2,-7/2})$ can be prepared with $\pi/2$ and $\pi$ pulses  using the coherent control developed for the implementation
of the octupole $^2S_{1/2}$ - $^2F_{7/2}$ atomic clock  \cite{HunLipTam14,GodNisJon14}. Alternatively, one can also implement $\pi/2$ and $\pi
$ pulses  by driving Raman transitions
via the  $^1$D$[5/2]_{5/2}$ state with 639 nm laser light
 (see Figure~\ref{fig:scheme}). For the odd $^{171}$\Yb~isotope which has a nuclear spin of $I=1/2$, one can prepare the target state $\ket{\Psi} = \ket{m_F=0,m_F = 0} + \ket{m_F=4,m_F = -4}(F=4)$ through the $^2\text{F}_{7/2}~ (F=3)$ state with radio-frequency pulses (see insert in Fig.~\ref{fig:scheme}).

The limits of the Lorentz-violating $c_{\mu\nu}$ parameters obtained from measurements with \Yb~ are expected to improve greatly over the current limits due to two main factors: sensitivity of the probe electronic state state and its lifetime. The experiment in Ref.~\cite{Pruttivarasin2014} with \Ca~was performed with a Ramsey duration of 100 ms, with the main limitation coming from the heating of the ion motion, which degrades the fidelity of the $\pi/2$ and $\pi$ pulses during the state readout. This problem can be overcome with sympathetic cooling
(see \cite{Pruttivarasin2014a}). However, the usable Ramsey duration is limited by the natural lifetime of the $^2$D$_{5/2}$ state approximately of 1.2~s to about 0.6~s. For the \Ba~ion, the lifetime of the $^2$D$_{5/2}$ state is approximately 30~s (Ref.~\cite{Iskrenova-Tchoukova2008}), which allows for Ramsey durations of $\sim$15~s, potentially improving the measurement precision by $\sqrt{15/0.6} \sim 5$ over \Ca.

Since the lifetime of the $^2\text{F}_{7/2}$ state of \Yb~ is in the range of several years\cite{Huntemann2012}, the Ramsey duration of the proposed experiment with \Yb~ is not limited by spontaneous decay during the measurement. Instead, the coherence of the target state in the decoherence-free subspace is likely to be limited by the stability of the magnetic field gradient in the vicinity of the ions. It has been shown that in a magnetically unshielded environment, ions can retain entanglement in the decoherence-free subspace for up to 30~s \cite{Haeffner2005a}. Hence, in a magnetically shielded trap, we can expect to suppress decoherence due to magnetic field noise such that Ramsey durations for much longer times are realistic.

The sensitivity of the frequency variation can be estimated from the quantum projection noise of the measurement:
\begin{equation}
\sigma_f = \frac{1}{4\pi A \sqrt{T}} \cdot \frac{1}{\sqrt{\tau}},
\end{equation}
where $A$ is the amplitude of the Ramsey oscillation signal, $T$ is the Ramsey duration and $\tau$ is the measurement time. Assuming that the target state $\ket{\Psi} = \frac{1}{\sqrt{2}}(\ket{1/2,-1/2} + \ket{7/2,-7/2})$ for $^2\text{F}_{7/2}$ is prepared through the creation of a mixed state as demonstrated in Refs.~\cite{Chwalla2007, Pruttivarasin2014}, then the amplitude of the Ramsey oscillation is $A = 0.5$. With the Ramsey duration of 60~s, the estimated sensitivity is $\sim 0.02/ \sqrt{\tau/ \text{Hz}}$. Together with the improved sensitivity of the atomic state to Lorentz-violation as given in Eq.~(\ref{eq:yb_sensitivity}), we estimate that the parameter $C_{0}^{(2)}$ will be bounded at the level of
 $1.7\times 10^{-19}/\sqrt{\tau/\text{sec}}$ for two ions,
 which is an improvement of $\sim2,400$ times
 over the measurement with \Ca (Ref.~\cite{Pruttivarasin2014}).
Use of the pure entangled two-ion states would improve these bounds by an additional factor of 2.

At this level of sensitivity, we expect that the $c_{JK}$ parameters will be bounded at the level of
$3\times10^{-22}$ and $1.5\times10^{-23}$
for a day-long and year-long measurement, respectively, for two entangled ions ($A=1$). The $c_{TJ}$ terms, which are sensitive to the velocity of the laboratory frame,  are suppressed by a factor of $10^{-4}$ due to the relatively small velocity of the Earth with respect to the cosmic microwave background as compared to the speed of light. With a year-long measurement, one expects to be sensitive to the $c_{TJ}$ parameters at the level of
 $1.5\times 10^{-19}$, a factor of 6,000
 improvement over the current best limits\cite{Hohensee2013c}. In the electron-photon sector, this would for the first time, probe all tensor Lorentz-violating elements below the ratio between the electroweak and Planck energy scales\cite{KosteleckyPottingPRD1995}.

Systematics which can directly affect the energy of our target states are magnetic and electric field fluctuations in the vicinity of the ions. Both fields can be measured directly using the ions themselves as a probe. One can then use the knowledge of these fields to correct for the effects of the differential linear Zeeman shift between the two ions due to a finite magnetic field gradient across the ion string, the quadratic Zeeman shift and the Stark shift from the oscillating electric field from the trap. A change in the environmental  temperature during the measurement can also affect the energy shift through the black-body radiation (BBR). The BBR shift between the two Zeeman states ($m_J=1/2$ and $m_J=7/2$) of the F$_{7/2}$ manifold depends on the anisotropy of the BBR source. By assuming that the spatial temperature variation of the environment is less than 0.1 K, we estimate that in the worst case scenario of a maximal anisotropy of 100\%, the differential BBR shift between the two Zeeman states is $\sim 1~\mu$Hz which affects the sensitivity of $c_{JK}$ at the $6\times10^{-24}$ level.

Our calculations also identified general rules for the enhancement of the reduced matrix elements of the $T^{(2)}$ operator. We find that the
only parameter that significantly affects the matrix elements leading to the tensor LV sensitivity is the deeper localization of the probe electron. We find $\langle \psi|r|\psi\rangle$ of $\sim$0.8~a.u or below for the corresponding electron to be a good indicator of  the large value of $T^{(2)}$ matrix element. This condition  is satisfied for the $4f$ hole states, such as Yb$^+$ state considered here, or for highly-charged ions with $nf$ valence electrons and degree of ionization $\sim 15$ \cite{clocks}.
Such systems with configurations containing  two $4f$ electrons of two $4f$ holes can have extra factor of two enhancement, and higher degree of ionization
leads to further increase in the $T^{(2)}$ matrix element values. Other lanthanide ions, for example the Tm$^+$ ion with $4f^{13}6s$ ground state, may provide other potential candidates for LV tests.
 We also considered the
  Th$^{3+}$ ion due to its $5f_{5/2}$ ground states and demonstrated laser cooling \cite{th}, but we find that the corresponding $T^{(2)}$ matrix element is more that factor of
 3 smaller than the one in Yb$^+$.  The $5f$ orbital  here is a valence orbital rather than a hole in the filled shell; moreover the $4f$ electron is generally deeper localized than the $5f$ one.

In summary, we identify \Yb ~ions as an excellent  system to test Lorentz symmetry in the electron-photon sector. We also develop general
rules for LV sensitivities in atomic systems. Systematic effects due to local magnetic field and electric field can be corrected for by using the ions themselves as an independent probes. The ultimate sensitivity of such experiments is likely to be limited by the differential BBR shift between the two Zeeman states, which we estimate to be approximately 10 times less than the projected sensitivity.

Due to increased sensitivity to LLI-violation and lifetime of the metastable F$_{7/2}$ state of \Yb, we estimate that experiments can reach the sensitivities of $1.5\times10^{-23}$ for the $c_{JK}$ parameters, more than $10^5$ times stringent than the current best limits\cite{Pruttivarasin2014}. Moreover, the projected sensitivity of the $c_{TJ}$ parameters will be at the level of $1.5\times 10^{-19}$, below the ratio between the electroweak and Planck energy scales.

It has been conjectured that Lorentz symmetry may be violated in string theories with the LV effects being suppressed
by some power of $m_{ew}/M_{pl}=2\times10^{-17}$, the ratio between the electroweak scale and the natural (Planck) energy scale for strings\cite{KosteleckyPottingPRD1995}.
If the Lorentz violation is not observed in the proposed here Yb$^+$ experiment, it will show that Lorentz violation in the photon-electron sector does not arise
at this first-order level in strings, or a cancellation is present between photon and electron LV effects to nullify the combined result.  With the proton and neutron tests already  over the O(1) bound \cite{Wolf2006,Allmendinger2013,Romalis2011},
such result would demonstrate that if physics at the Planck scale violates Lorentz invariance, it is more than minimally suppressed at low energies for normal matter.\\

 \noindent \textbf{Acknowledgements}\\
 \noindent M. S. S. thanks the School
of Physics at UNSW, Sydney, Australia  for hospitality and
acknowledges support from the Gordon Godfrey
Fellowship program, UNSW.
This work was supported by the NSF CAREER
program grant \# PHY 0955650, NSF grant  \# PHY
1404156, the Australian Research Council and was performed under the auspices of the U.S. Department of
Energy by Lawrence Livermore National Laboratory under Contract
DE-AC52-07NA27344. T.P. is supported by RIKEN’s Foreign Postdoc Researcher program.
Free clip art from http://www.clker.com/clipart-globe-2.html is used in Fig. 2. \\

\noindent \textbf{Contributions}\\
\noindent V.A.D. had the idea to use Yb$^+$ and identified the enhancement to LV violation. V.A.D., V.V.F., M.S.S., and  S.G.P.
carried out atomic calculations. T.P., M.A.H. and H.H. worked out the experimental scheme considerations and projected
 LV limits. All authors contributed to the discussions of the results and manuscript.\\

\noindent \textbf{Competing financial interests}\\
\noindent The authors declare no competing financial interests.\\

\noindent \textbf{Corresponding author}\\
\noindent Correspondence to: M.S. Safronova.

\noindent \textbf{METHODS}\\

\noindent The calculations
of the energy shift due to LLI violation  reduces to the calculation of the expectation value of the Hamiltonian in Eq. (\ref{eq1}).
 The matrix element of $\langle J m_J|\mathbf{p}^{2}-3p_{z}^{2}| Jm_J \rangle$ is expressed through
the reduced matrix element of the $T^{(2)}$ operator using the Wigner-Eckart theorem
\begin{equation*}
\langle Jm_J|T^{(2)}_{0}|Jm_J\rangle=(-1)^{J-m_J}\left(
\begin{array}
[c]{ccc}%
J & 2 & J\\
-m_J & 0 & m_J
\end{array}
\right)  \langle J||T^{(2)}||J\rangle.
\end{equation*}
Using the algebraic expression for the
$3j$-symbol, we arrive at the following expression for
the matrix element of  $\langle Jm_J|\mathbf{p}^{2}-3p_{z}%
^{2}|Jm_J \rangle$ operator:
\begin{eqnarray}
\label{eq10}
\langle Jm_J|T^{(2)}_{0}|Jm_J \rangle &=& \frac{-J\left(  J+1\right)  +3m_J^{2}}{\sqrt{\left(
2J+3\right)  \left(  J+1\right)  \left(  2J+1\right)  J\left(  2J-1\right)}} \,\nonumber \\ &\times&
\langle J||T^{(2)}||J \rangle.
\end{eqnarray}

\begin{table}       [ht]
\caption{\label{tab2} Reduced matrix elements of $T^{(2)}$ operator in
 Ca$^+$, Ba$^+$, Yb$^+$ ions in atomic units. Ca$^+$ values are from Ref.~\cite{Pruttivarasin2014}. Method of calculations is listed in the column ``Method''. LCCSD is linearised coupled-cluster method with single and double excitations,
 DF is Dirac-Fock method, RPA is random-phase approximation, and CI is configuration iteration. We list the included configurations for the 15-electron CI calculation.   }
\begin{ruledtabular}
\begin{tabular}{lllc}
Ion    & State &    Method & $\langle J||T^{(2)}||J \rangle$       \\ \hline
Ca$^+$ & $3d$~$^2D_{3/2}$ & LCCSD &  7.09(12) \\
       & $3d$~$^2D_{5/2}$ & LCCSD & 9.25(15)   \\ [0.4pc]
Ba$^+$ & $5d$~$^2D_{3/2}$ & LCCSD &6.83   \\
       & $5d$~$^2D_{5/2}$ & LCCSD &8.65   \\ [0.4pc]
Yb$^+$ & $4f^{14} 5d$~$^2D_{3/2}$ & LCCSD & 9.96 \\ [0.4pc]
       & $4f^{14} 5d$~$^2D_{5/2}$ &  DF           & 7.23 \\
       &                           & CI $7sp6df5g$ & 11.6 \\
       &                          & LCCSD         & 12.08   \\ [0.4pc]
       & $4f^{13} 6s^2$~$^2F_{7/2}$ &  DF   &   -145     \\
       &                          & CI~$6sp5df$ & -143.8 \\
       &                           & CI+RPA ~$6sp5df$ & -135.1 \\
       &                             & CI+RPA $7sp6df5g$ &  -135.6 \\
       &                             & CI+RPA $8sp6df5g$ &  -135.2 \\
\end{tabular}
\end{ruledtabular}
\end{table}
 The all-order method gave very accurate
values of the $3d_J$ lifetimes~\cite{KreBecLan05} and quadrupole moments~\cite{JiaAroSaf08} in Ca$^+$.
In the  all-order method, single, double, and partial triple excitations of Dirac-Hartree-Fock wave functions are included to all orders
of perturbation theory. We refer the reader to review Ref.~\cite{SafJoh08} for the description of the all-order method and its applications. The all-order results are accurate to about 1.5\% \cite{Pruttivarasin2014}.
 The calculations were carried out with both nonrelativistic and relativistic  operators;
the differences were found to be negligible at the present level of accuracy.
The results are listed in Table~\ref{tab2}.

The all-order method was designed to work for monovalent systems and is not applicable to the calculation of $4f^{13} 6s^2$~$^2\text{F}_{7/2}$ properties due
 to the electronic configuration of this state that has a hole in the $4f$ shell.
We used both a one-electron Dirac-Fock calculation and a large-scale configuration interaction calculation for this state and find excellent agreement of both approaches.
The 15-electron configuration interaction calculation  follows the approach described  in Ref.~\cite{PorSafKoz12Ybp}.
Briefly, we start from the solution of the Dirac-Fock equations and
 carry out the initial self-consistency procedure for the [1$s^2$,...,4$f^{14}$, 6$p$]
configuration. The $6s$ orbital was constructed for the
$4f^{13} 6s^2$ configuration, and the
$5d_{3/2,5/2}$ orbitals were constructed for the $4f^{13} 5d6s$
configuration.
The basis set used in the CI calculations included virtual orbitals
up to $8s$, $8p$, $7d$, $7f$, and $5g$. The virtual orbitals were
constructed as described in Refs.~\cite{Bog91,KozPorFla96}. The configuration space was formed by
allowing single and double excitations for the odd-parity
states from the $4f^{14} 6p$, $4f^{13} 6s^2$ and
$4f^{13} 5d6s$ configurations to virtual orbitals of the basis set listed above.

We have verified the convergence of the CI by carrying out three calculations with an
increasing set of configurations functions: (1) including single and
double (SD) excitations to the $6s$, $6p$, $5d$, and
$5f$ orbitals (we designate it [$6sp5df$]), (2) adding excitations to the $7s$, $7p$, and
$6d$, $6f$ and $5g$ orbitals [$7sp6df5g$], and (3) also adding $8s$, $8p$, $7d$, and $7f$ orbitals
[$8sp7df5g$]. The last configuration set results in rather lengthy calculations with $\sim 2\,300\,000$ determinants.

As demonstrated in Table~\ref{tab2}, the number of included configurations has only a negligible effect on the
$T^{(2)}$ operator of the $4f^{13} 6s^2$~$^2F_{7/2}$  state. The CI number is in agreement with simple DF
calculation. Inclusion of the random-phase approximation changes this value by 6\%.

\end{document}